\documentclass{aa}  

\usepackage{graphicx}  
\usepackage{txfonts}
\usepackage{lipsum}
\usepackage{natbib}
\usepackage{subcaption}        
\usepackage{lscape}                      
\usepackage{placeins}           

\usepackage[colorlinks=true,
    linkcolor=blue,
    citecolor=blue,
    urlcolor=blue]{hyperref}

\usepackage[dvipsnames]{xcolor}
\usepackage[]{ulem}
\def\msun{\rm\, M_\odot}
\usepackage{verbatim}

\defcitealias{Lupi2024}{L24}
\defcitealias{Maiolino2025}{M25}

\titlerunning{Highly accreting X-ray weak AGNs at cosmic dawn}
\authorrunning{Trinca et al.}

\begin{document}

   \title{You can’t see me: Super-Eddington growth hindering X-ray detection in high-z broad-line active galactic nuclei}

   \subtitle{}

   \author{
       Alessandro Trinca\inst{1,2,3} \fnmsep\thanks{atrinca@roe.ac.uk} \and
       Alessandro Lupi\inst{3,4,5} \and
       Francesco Haardt\inst{6,4,7}
       \fnmsep\thanks{on leave from CLAP, Insubria, Italy}\and
       Piero Madau \inst{8,9}
        }

   \institute{INAF, Osservatorio Astronomico di Roma, Via Frascati 33, 00078 Monte Porzio Catone, Italy
        \and Institute for Astronomy, University of Edinburgh, Royal Observatory, Blackford Hill, Edinburgh EH9 3HJ, UK
        \and
            Como Lake Center for Astrophysics, DiSAT, Universit\`a degli Studi dell'Insubria,  via Valleggio 11, 22100, Como, Italy
        \and
            INFN, Sezione di Milano-Bicocca, Piazza della Scienza 3, I-20126 Milano, Italy
         \and
         INAF, Osservatorio Astronomico di Bologna, Via Gobetti 93/3, I-40129 Bologna, Italy  
        \and
         Center for Astrophysics and Space Science (CASS), New York University Abu Dhabi, PO Box 129188, Abu Dhabi, UAE
        \and
        INAF, Osservatorio Astronomico di Brera, Via E. Bianchi 46, I-23807 Merate, Italy
        \and
         Department of Astronomy \& Astrophysics, University of California Santa Cruz, 1156 High Street, Santa Cruz, CA 95064, USA
         \and
          Dipartimento di Fisica "G. Occhialini", Università degli Studi di Milano-Bicocca, Piazza della Scienza 3, I-20126 Milano, Italy
  }

\date{Received 20 February 2026 / Accepted 24 April 2026}

\abstract{
 We revisit black hole mass estimates for high-redshift broad-line active galactic nuclei (AGNs) discovered with JWST by jointly analysing their broad emission lines and their systematic non-detections in deep \textit{Chandra} imaging. Building upon a self-shadowed, super-Eddington accretion framework in which the corona undergoes efficient radiative over-cooling, we couple funnel-dependent Comptonisation physics with slim-disc spectral models and explore the resulting parameter space through a full Markov Chain Monte Carlo inference. Using a recently compiled sample of JWST high-redshift broad-line AGNs, we show that the observed X-ray weakness -- manifested as extreme bolometric corrections, suppressed 2--10 keV luminosities, and non-detections in the 0.5--5 keV \textit{Chandra} band -- naturally arises when the corona is confined and radiatively over-cooled inside a narrow super-Eddington funnel.
 The combined broad line+X-ray analysis yields strongly bimodal posteriors: either very massive, very low-Eddington black holes (physically disfavoured), or a population of low-mass ($\sim 10^{6}$--$10^{7}\,M_{\odot}$) black holes accreting at $f_{\rm Edd} \gg 1$. The latter solution is strongly preferred for nearly all objects and returns masses consistent with, or lower than, local $M_{\rm BH}$--$M_{\star}$ relations, mitigating the extreme mass ratios implied by single-epoch virial estimators. The predicted intrinsic spectra are redder and exhibit reduced hard-X-ray output but higher bolometric luminosities, implying bolometric corrections larger than those typical of the local AGN population, yet consistent with low-redshift highly accreting counterparts. These results support a picture in which many JWST broad-line AGNs are powered by rapidly growing, super-Eddington black holes whose suppressed coronal emission and self-shadowed broad-line region geometry combine to mimic overmassive black holes at $z \gtrsim 6$.
}


   \maketitle

\section{Introduction}

Massive black holes (MBHs) are ubiquitously found at the centre of galaxies up to redshifts well above $6$ \citep[e.g.][]{fan2003,mortlock2011,banados2018,fan2022,Maiolino2024bhs}, with masses in the range of $\sim 10^5$--$10^{10}\,\msun$. The observational evidence of their presence comes from the detection of electromagnetic emission produced in their vicinity by gas accretion, which makes them shine as active galactic nuclei (AGNs). Exploiting this information to infer MBH properties is, however, extremely challenging because of the complex structure of accretion flows \citep{urry1995}. 

In the local Universe, reverberation mapping campaigns have provided unique information about the properties of the so-called broad-line region (BLR), a region around the MBH characterised by fast gas motions (up to several thousand km s$^{-1}$) that broaden the gas emission at specific frequencies (e.g. the Balmer series or the Mg\,II and C\,IV lines). Under the assumption of virial equilibrium for the BLR gas, a correlation between black hole (BH) mass and the emission properties of these lines has been calibrated \citep[e.g.][]{Vestergaard2009,bentz2013,reines2015,greene2020}, providing a simple but effective method of inferring BH masses from single-epoch spectra. Over the years, this method has been applied to a wide variety of sources, including the highest-redshift objects observed with ALMA and JWST. Interestingly, most of these high-redshift objects appear to be more massive than their local counterparts when normalised to the masses of their host galaxies \citep{farina2022,Maiolino2024bhs,yue2024,Juodzbalis2024}. Such large mass ratios challenge most theoretical models for the formation and growth of MBHs and require favourable conditions such as a primordial origin \citep{ziparo2022,Juodzbalis2025} or initially heavy seeds ($10^4$--$10^5\,\msun$, see e.g. \citealt{begelman2006formation,volonteri2010,latif2016b}) growing continuously at the Eddington limit. Another challenge arises from their large observed abundance \citep[e.g.][]{harikane2023,Maiolino2024bhs,Greene2024}, which implies a formation efficiency much higher than found in most theoretical models. Several studies have shown that this issue can be alleviated if accretion above the Eddington limit is allowed \citep[e.g.][]{madau2014a,Lupi2016,  pezzulli2016,regan2019,massonneau2023,lupi2024SE,Shi2024,Trinca2024b,Husko2025,Quadri2025}, which can compensate for stunted growth in low-mass galaxies \citep[see e.g.][]{angles-alcazar2017}.
A further concern about this population of high-redshift MBHs is the weak or absent X-ray emission \citep{Maiolino2025}. Hard X-rays are commonly produced by the AGN corona, a hot ionised medium that scatters UV photons from the inner accretion disc into the X-ray band via Comptonisation. The lack of detectable X-ray emission casts doubts on the nature of these broad-line emitters and has renewed interest in alternative explanations, including the possibility of super-Eddington accretion.

Although rarely considered for AGNs, super-Eddington accretion is common in ultra-luminous X-ray sources and tidal disruption events. In the past decade, the first evidence of super-Eddington accretion around local MBHs has been provided by the SEAMBH and SDSS-RM campaigns \citep{Du2018,MartinezAldama2019,Grier2017,Du2019}. Motivated by these results, many works have proposed super-Eddington accretion as a plausible explanation for the peculiar properties of high-redshift MBHs, especially the so-called ``little red dots'' \citep[LRDs,][]{Matthee2024}. These include their weak X-rays \citep{Lambrides2024,Madau2024} and the lack of detectable variability \citep{Secunda2025}. Several studies have also suggested that the gas distribution around MBHs during super-Eddington accretion phases may strongly affect the electromagnetic emission, introducing biases in the black hole masses inferred using locally calibrated scaling relations \citep[e.g.][]{reines2015,Vestergaard2009}. In particular, the broad-line profile might be altered by several physical mechanisms, including radiation beaming \citep{king2024,Madau2026}, angle-dependent shadowing of BLR clouds \citep{Lupi2024}, scattering processes, \citep{Rusakov2025,Chang2026}, and small-scale inflows or outflows that significantly affect the observed optical line velocities \citep[see e.g.][]{Torralba2026, Deugenio2026}. Direct evidence of such a bias has recently been provided by resolved BLR observations at intermediate redshift by the GRAVITY+ collaboration \citep{Abuter2024,Dayem2025}, which found highly accreting BHs with masses up to an order of magnitude smaller than those inferred from single-epoch methods.

Motivated by these developments and by our previous analysis of how the single-epoch method biases BH mass estimates upwards in super-Eddington systems, we extend our study to the apparent X-ray weakness of these sources. We re-analyse the 14 objects examined in \citet[][L24 hereafter]{Lupi2024} and included in \citet[][M25 hereafter]{Maiolino2025}. Starting from the \citet{Madau2024} model, in which the AGN corona in super-Eddington accretion flows is confined and cooled inside the central funnel of a thick accretion disc, we aim to reassess BH masses and infer accretion rates and possibly BH spins under the assumption that the observed lack of X-rays is due to an intrinsically steeper X-ray spectrum, as opposed to extreme gas column densities.

The manuscript is organised as follows. In Section~\ref{sec:methods} we describe our procedure to estimate BH masses using the constraints from X-ray emission. In Section~\ref{sec:results} we present our results. In Section~\ref{sec:conclusions} we discuss caveats in the analysis and summarise our conclusions.

\section{Methods}
\label{sec:methods}
For our purposes, in this work we couple the model for the X-ray weakness of super-Eddington flows by \citet{Madau2024} with the slim-disc spectrum by \citet{Kubota2019}, and incorporate them into a Markov chain Monte Carlo (MCMC) analysis tool as implemented in the \textsc{emcee} package \citep{foremanmackey2013} to more accurately infer BH masses, accretion rates, and spins.

First, we computed the effective hard X-ray temperature and slope as a function of the half-opening angle of the funnel following \citet{Madau2024}, assuming $f_c=1$ and a single scattering probability in crossing the corona of 0.3. The results of these calculations, described in detail in Appendix~\ref{app:corona}, are shown in Fig.~\ref{fig:gamma}. We then connected the half-opening angle of the funnel to the MBH accretion rate using Eq.~11 in \citet{Wang2014}, adjusted to ensure continuity across the entire range,
\begin{equation}
    \theta_{\rm fun}=
    \left\{
    \begin{array}{lc}
    83.1^\circ & f_{\rm Edd}<10\\[4pt]
    60^\circ - 33^\circ \log(f_{\rm Edd}/50) & 10 \leq f_{\rm Edd} < 76.5 \\[4pt]
    52^\circ - 12^\circ \log(f_{\rm Edd}/100) & f_{\rm Edd} \geq 76.5,\\    
    \end{array}
    \right.
\end{equation}
where $f_{\rm Edd}=\dot{M}_{\rm BH}c^2/L_{\rm Edd}$, $\dot{M}_{\rm BH}$ is the MBH accretion rate, $c$ is the speed of light, $L_{\rm Edd}=1.26\times 10^{38}(M_{\rm BH}/{\msun})\rm\, erg\, s^{-1}$ is the Eddington luminosity, and $M_{\rm BH}$ is the BH mass.

\begin{figure}
    \centering
    \includegraphics[width=\columnwidth]{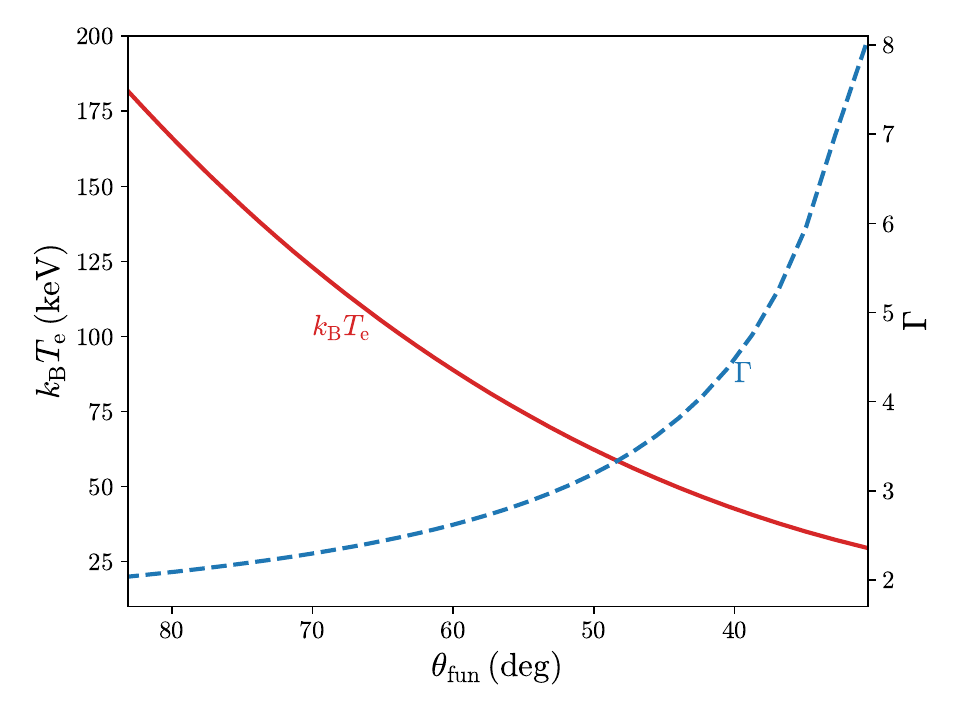}
    \caption{Effective electron temperature (red solid line) and X-ray photon index $\Gamma$ (dashed blue line) as a function of the half-opening angle of the accretion disc funnel, as predicted by the model of \citet{Madau2024}. The model accounts for coronal cooling due to inverse-Compton scattering of disc photons, which becomes increasingly efficient at high accretion rates, where the disc is geometrically thicker and the funnel correspondingly narrower. In this regime, the enhanced soft-photon flux leads to stronger cooling of the coronal plasma, resulting in progressively steeper hard X-ray spectra with increasing Eddington ratios. Details of the calculations are provided in Appendix~\ref{app:corona}.}
    \label{fig:gamma}
\end{figure}

We then built a library of AGN spectra using the \textit{agnslim} model \citep{Kubota2019} in \textsc{xspec}, as a function of the BH mass $M_{\rm BH}$, the MBH spin $a_\bullet$, and the thin-disc Eddington ratio $f_{\rm Edd, thin}=\eta_{\rm thin}(a_\bullet)\dot{M}_{\rm BH}c^2/L_{\rm Edd} = \eta_{\rm thin}(a_\bullet)f_{\rm Edd}$, with $\eta_{\rm thin}(a_\bullet)$ the spin-dependent radiative efficiency for a \citet{shakura1973} accretion disc. Unlike in our previous work, in which the other parameters of the \textit{agnslim} model were left at their default values, here we also include variations in the slope (\textit{Gamma\_hot}) and effective electron temperature (\textit{kTe\_hot}) of the hot Comptonisation region of the model, with \textit{Gamma\_hot} in the range 2–8 and \textit{kTe\_hot} in the range 30–200~keV.

We sampled 7,500 different combinations, with 25 logarithmically spaced BH masses between $10^5$ and $10^{10}\,\msun$, 30 logarithmically spaced values of $f_{\rm Edd,thin}$ in the range $10^{-3}$--$10^3$, and 10 linearly spaced values of $a_\bullet$ between 0 and 0.998. Each spectrum covers the energy range $0.01$~eV--$200$~keV, corresponding to a wavelength range $1.2$\,\AA\--$24.8\,\mu$m, in 40,000 logarithmically spaced bins.
After the spectra have been generated, for each combination we tabulated the bolometric luminosity, $L_{\rm bol}$, the rest-frame X-ray luminosity in the 2–10~keV band, $L_{\rm 2-10}$, and the corresponding observer-frame luminosity, $L_{\rm X}$, in the 0.5–5~keV band observable by \textit{Chandra}, considering 30 redshift bins between $z=4$ and $z=7$, in addition to the quantities employed in \citetalias{Lupi2024}. For consistency with \citet{Kubota2019}, we normalised the bolometric luminosity of each spectrum to the value estimated from the numerical integration of the slim-disc solution by \citet{Sadowski2011}. With these tables, we constructed a multidimensional interpolation that was used in the likelihood evaluation.

For our analysis, we considered as our observational sample a subset of the sources analysed by \citetalias{Maiolino2025}, in particular those for which a BH mass estimate based on our previous analysis of the BLR emission from \citetalias{Lupi2024} was available. We defined the likelihood, $\mathcal{L}$, for the MCMC as
\begin{equation}
    \ln \mathcal{L} = -\frac{1}{2} \sum_i \left[ \frac{(Y_i-\bar{Y}_i)^2}{s_i^2} + \ln (2\pi s_i^2)\right] - \infty\,\mathrm{H}(L_{\rm X}-L_{\rm X,lim}),
\end{equation}
where $Y_i$ can be (i) the bolometric luminosity or (ii) the observed luminosity and full width at half maximum (FWHM) of the H$\alpha$ broad line of the source, $\bar{Y}_i$ is the value predicted by our model, $s_i$ is the uncertainty in the observed data (assumed Gaussian), and $H$ is the Heaviside step function, evaluated using the difference between the X-ray luminosity from the model and the minimum observable luminosity in the 0.5–5~keV \textit{Chandra} band. The Heaviside function was introduced to ensure that every model predicting a detectable X-ray flux was automatically rejected.

The parameters of our model that we aim to constrain are $M_{\rm BH}$, $L_{\rm thin}/L_{\rm Edd}$, and $a_{\rm BH}$. As priors, we assumed a log-flat distribution for $M_{\rm BH}$ and $L_{\rm thin}/L_{\rm Edd}$ over the intervals $[5,10]$ and $[-3,3]$, respectively, and a uniform distribution for $a_{\rm BH}$ between 0 and 0.998. 
We ran the MCMC for 10{,}000 steps employing 32 walkers.\footnote{The number of steps chosen corresponds to about 100 autocorrelation time-scales, which is sufficient to guarantee robust optimisation.}

\section{Results}
\label{sec:results}

\subsection{Bolometric correction comparison}
Before running the MCMC, we first performed a benchmark test of our mock AGN spectra, which follow the super-Eddington flow prescriptions of \citet{Madau2024} and include the associated changes in the X-ray band. In particular, we investigated the range of bolometric corrections expected for the high-redshift AGN population by adopting the values of $M_{\rm BH}$ and $L_{\rm bol}$ reported by \citetalias{Maiolino2025}, which are inferred from standard scaling relations \citep{Stern2012}. From these estimates, we evaluated $K_{\rm X} = L_{\rm bol}/L_{\rm X}$ using our \textsc{xspec} tables for different values of $a_\bullet$. 

\begin{figure}
\centering
\includegraphics[width=\hsize]{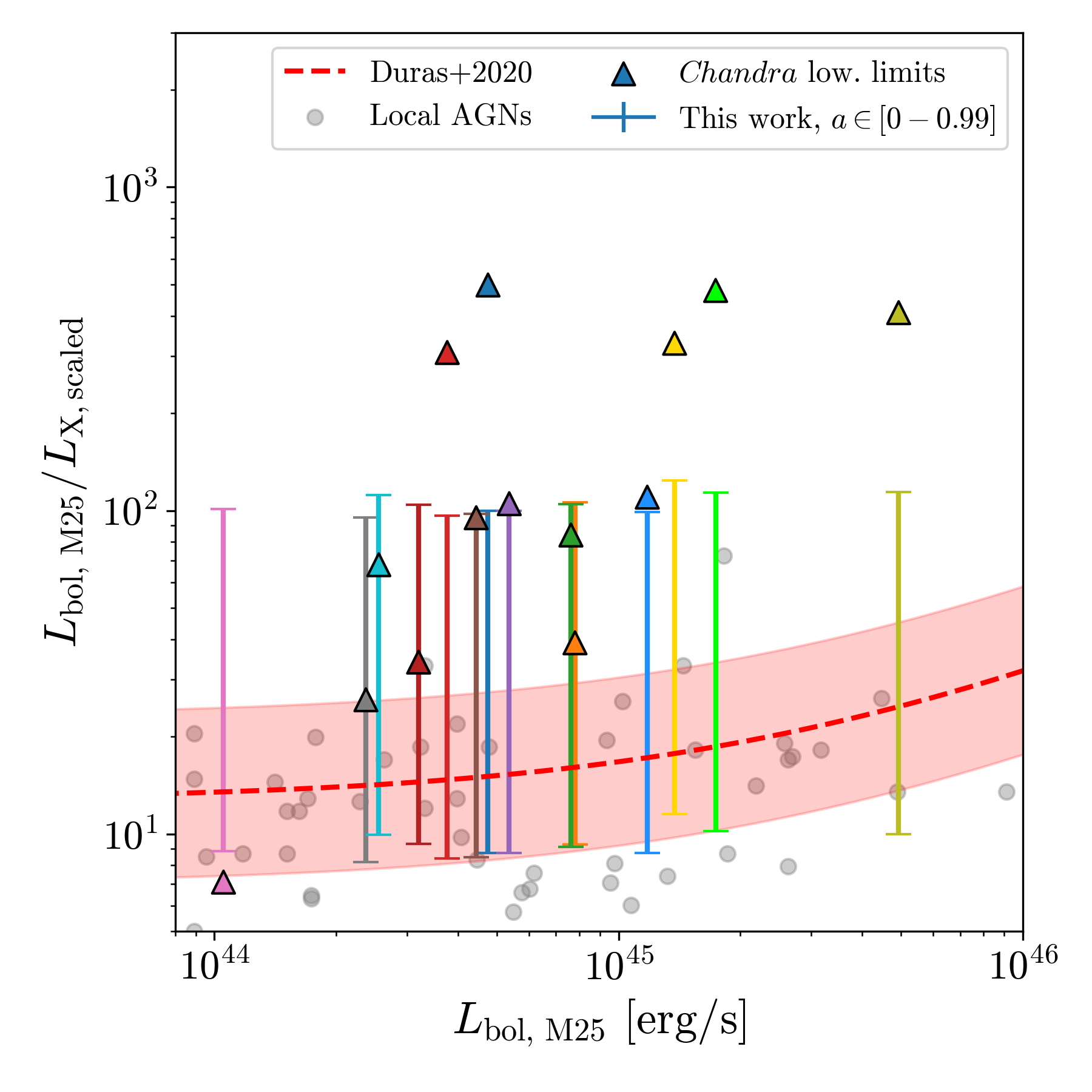}
\caption{Comparison between the expected $K_{\rm X}$ for our sample of high-redshift AGNs (assuming the $M_{\rm BH}$ and $f_{\rm Edd}$ estimates from \citetalias{Maiolino2025}) for different spin values, and the observed lower limits from the X-ray non-detection, shown as triangles. Our models are shown with solid vertical bars spanning values expected for different spins, ranging between $a_\bullet=0$ (upper bounds, implying higher bolometric corrections) and $a_\bullet=0.99$ (lower bounds). The local scaling relation is shown as a red line, with the shaded area corresponding to a 1$\sigma$ uncertainty ($\pm 0.26 ~\rm dex$), and the local observations by \citet{duras2020} as grey points. Colour bars extending above the corresponding observed lower limits (shown in the same colour) indicate the existence of spin values consistent with the current \textit{Chandra} non-detection.}
\label{fig:Kx_M25}
\end{figure}

\begin{figure}
\centering
\includegraphics[width=\hsize]{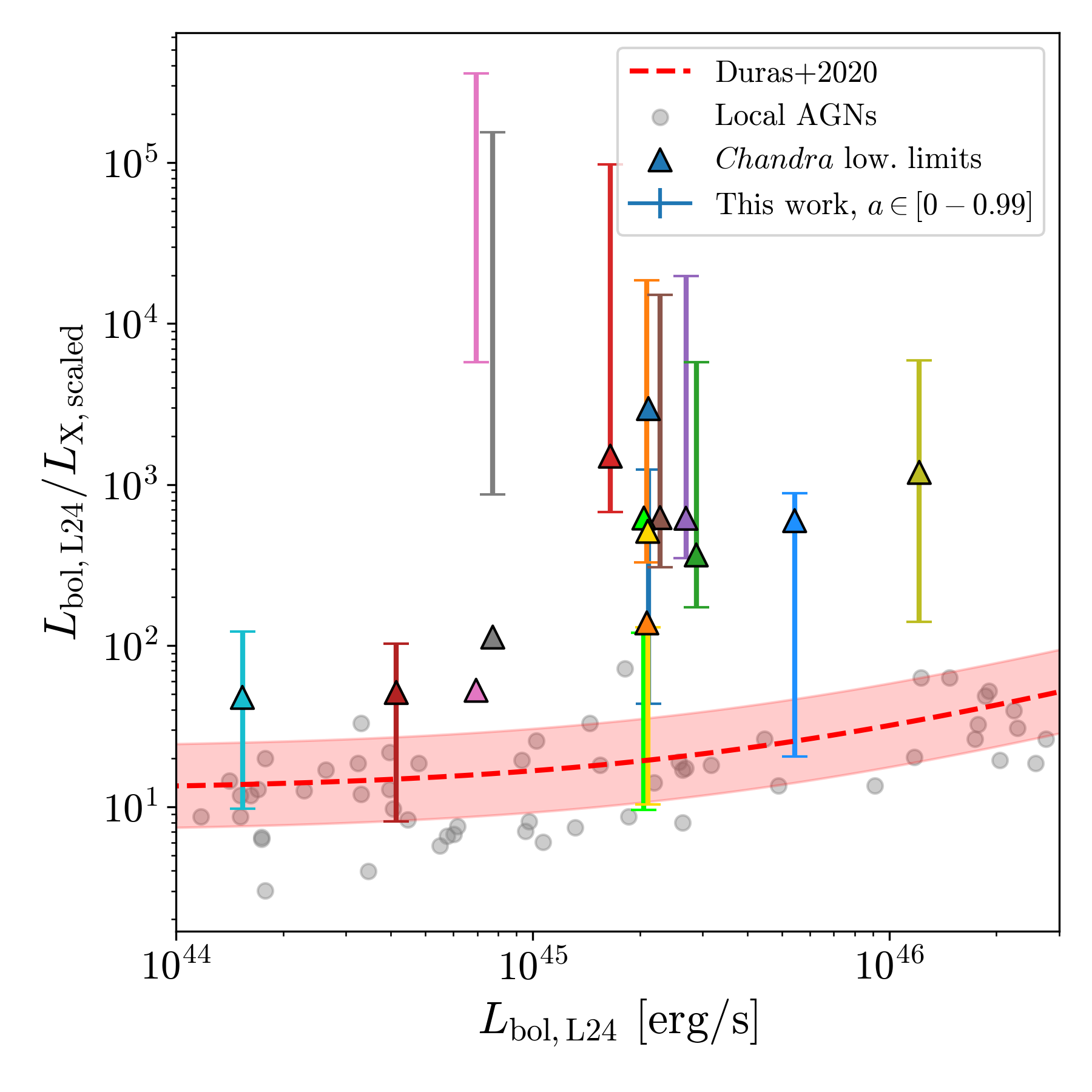}
\caption{Same as Fig. \ref{fig:Kx_M25} but assuming BH masses and accretion rates from \citet{Lupi2024}. We selected here the combination of $M_{BH}, f_{Edd}, a_\bullet$ showing the highest $K_{X}$ correction among the ten solutions with the highest probability over the 32k samples. Bolometric corrections are corrected according to the predicted bolometric luminosity of each source. Nearly all systems allow for spin configurations consistent with the current X-ray non-detection.}
\label{fig:Kx_L24}
\end{figure}

In Fig.~\ref{fig:Kx_M25}, we show the lower limit on $K_{\rm X}$ provided by \citetalias{Maiolino2025} for each source, compared to the corresponding range of $K_{\rm X}$ values expected for different spins between $a_\bullet = [0-0.99]$.\footnote{Note that the values of $L_{\rm bol}$ reported here may differ slightly from those in \citetalias{Maiolino2025}, since they are consistently estimated from the \textsc{xspec} tables based on the inferred $M_{\rm BH}$ and $f_{\rm Edd}$.} Low spin values, which imply higher bolometric corrections, represent the upper bound of the range, and maximal spin values the lower bound. Bars extending above the corresponding lower limits indicate spin configurations consistent with current \textit{Chandra} non-detections. For reference, we also show the local scaling relation and the compilation of data from \citet{duras2020}.
From this comparison, we see that, even without modifying the X-ray spectral slope according to the super-Eddington prescriptions, adopting a more conservative choice of $\Gamma \approx 2$ (instead of $\Gamma=1.7$ assumed in \citetalias{Maiolino2025}) for low-accretion-rate sources -- as predicted in Fig. \ref{fig:gamma} for a large opening angle, $\theta \rightarrow 90^\circ$ -- is sufficient to explain the X-ray non-detections with standard AGN spectra for at least a subset of spin values in $\sim 40 \%$ of sources. Only the sources exhibiting very high bolometric corrections appear clearly incompatible with our models when assuming the literature estimates for $\rm M_{BH}$ and $\rm L_{bol}$.
We then repeated this analysis by adopting instead the BH masses and bolometric luminosities reported in \citetalias{Lupi2024}, allowing for the possibility of super-Eddington accretion. The results are shown in Fig.~\ref{fig:Kx_L24}. To remain consistent with the original analysis in \citetalias{Lupi2024}, we retrieved the values of $M_{\rm BH}$ and $f_{\rm Edd,thin}$ for the ten highest-probability solutions among the 32,000 samples of the original MCMC run, and we report the range of values of $K_{\rm X}$ for different spins for the solution with the highest bolometric correction. We find that, based solely on the \citetalias{Lupi2024} estimates, the predicted range of bolometric corrections extends above the corresponding empirical lower limit for nearly all systems, implying that there exists a range of spin configurations that is compatible with the current X-ray non-detections.

Only two sources in the sample show a range of $K_{\rm X}$ values that is entirely below the corresponding lower limit when using the \citetalias{Lupi2024} estimates for $M_{\rm BH}$ and the accretion rate. As discussed below, our subsequent analysis finds for these two sources a preference for higher $L_{\rm bol}$ values than in \citetalias{Lupi2024}, which naturally explains their non-detections.

\subsection{MCMC analysis}
We now extend our analysis and infer the best-fitting model parameters using the MCMC procedure described in Section~\ref{sec:methods}. We consider two different cases: one in which the bolometric luminosity is fixed to the value estimated by \citetalias{Maiolino2025} from standard correlations, and another in which we instead constrain the broad-line properties as in \citetalias{Lupi2024}, together with the X-ray non-detection of the target sources. 

\begin{figure}
\centering
\includegraphics[width=0.99\columnwidth]{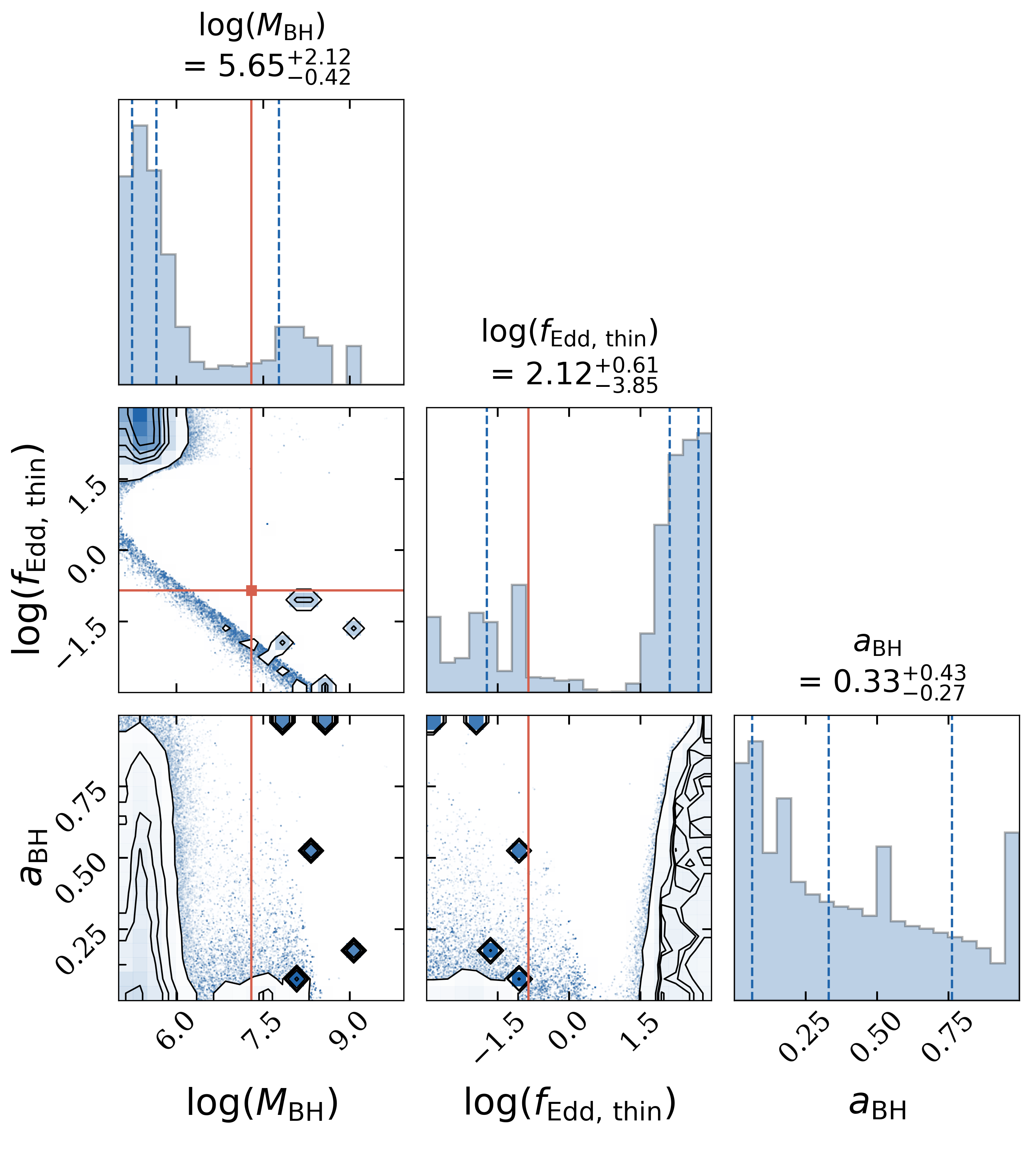}
\caption{Example of a corner plot from the MCMC X-ray analysis for the \textsc{GS8083} source. The orange lines and square marker show the reference values of $M_{\rm BH}$ and $f_{\rm Edd}$ from \citetalias{Maiolino2025}. The numerical values reported above each marginalised distribution correspond to the medians of the respective posteriors, while the $[0.16, 0.5, 0.84]$ quantiles are shown as dashed blue lines. A bimodal structure is clearly visible in the posterior distributions of BH mass and accretion rates, with current literature estimates lying between the two solutions. Spin values are poorly constrained across the full set of MCMC samples.}
\label{fig:corner}
\end{figure}

\subsubsection{Mass estimates from the X-ray weakness alone}
\label{sec:MCMC_xray}

As an example of the results of our analysis, Fig.~\ref{fig:corner} shows the corner plot for the source \textsc{GS8083}, where $M_{\rm BH}$ and $f_{\rm Edd}$ estimates provided in \citetalias{Maiolino2025} are also reported. 
We observe that, regardless of the assumption on $L_{\rm bol}$, the posterior distributions of both the BH mass and Eddington ratio are bimodal. This behaviour is not unique to this single candidate, but is observed across all the systems we analysed.

This result implies that the observational constraints considered -- namely, the expected $L_{\rm bol}$ and the X-ray weakness -- can be interpreted in two contrasting ways: (i) the lack of X-ray emission is due to the increased steepness of the X-ray spectrum when the accretion rate exceeds the Eddington limit, so that we are observing a highly accreting, low-mass BH; or (ii) a very massive BH is accreting well below Eddington, close to the advection-dominated accretion flow (ADAF) regime, where the intrinsic X-ray luminosity is low despite a relatively shallow spectrum. Interestingly, the mass estimate by \citetalias{Maiolino2025} lies between the two values preferred by our MCMC analysis, as shown in Fig. \ref{fig:corner}. 

\begin{figure*}
\centering
\includegraphics[width=\hsize]{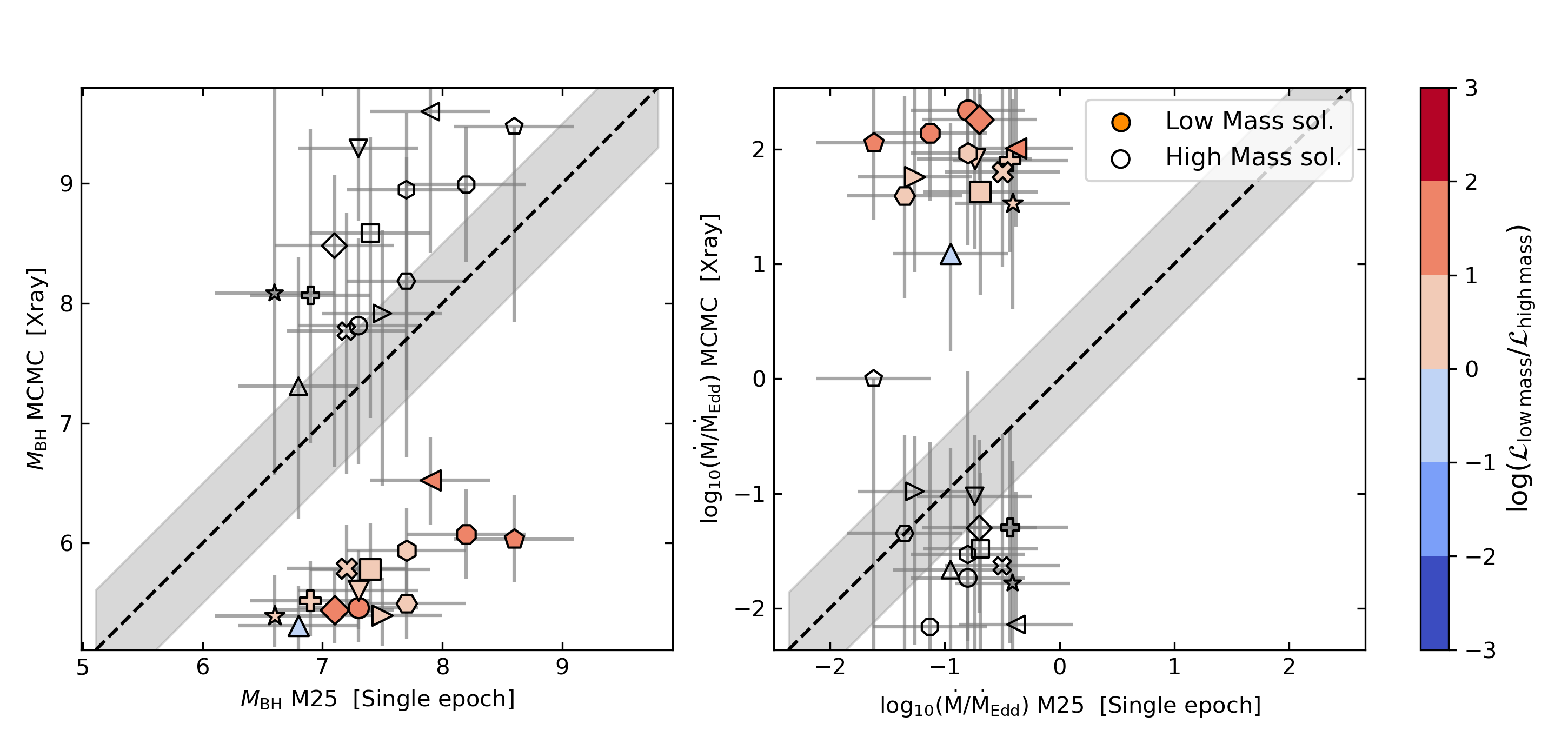}
\caption{Comparison between the best-fit values of $M_{\rm BH}$ (left panel) and $\dot{M}/\dot{M}_{\rm Edd}$ (right panel) derived from the MCMC analysis (which includes constraints on the X-ray emission) and the estimates from \citetalias{Maiolino2025} obtained using the single-epoch method. Given the bimodal posterior distributions found for all objects, the Monte Carlo samples are divided into two groups: low-mass, highly accreting systems (filled data points) and high-mass, almost inactive black holes (empty data points). Each pair of solutions for a given object is represented by a different marker. Data points for the low mass solution are colour-coded according to the ratio between the mean likelihoods obtained for the low-mass and high-mass solutions, where high values (red colours) suggest a preference for the former. The shaded grey regions indicate a deviation of 0.5~dex from the 1:1 relation.}
\label{fig:mcmc_vs_M25}
\end{figure*}

Because of this bimodality, the median values for the different parameters typically obtained through Bayesian analysis cannot be taken as representative of the actual distribution. For this reason, we opted to split the samples from the MCMC into two populations. To optimally separate the different solutions, we assumed $f_{\rm Edd,thin} = 1$ as a threshold. The results after this splitting are shown in Fig.~\ref{fig:mcmc_vs_M25}, where each source is represented by a characteristic marker. In particular, we report the $M_{\rm BH}$ and $\dot{M}/\dot{M}_{\rm Edd}$ \footnote{where $\dot{M}_{\rm Edd} \equiv 16 ~L_{\rm Edd} / c^2$} values obtained from the MCMC for the two different solutions, plotted as a function of the estimates from \citetalias{Maiolino2025}. The shaded region indicates the 1:1 relation with a $0.5 ~\rm dex$ dispersion, where systems consistent with literature estimates are expected to lie. We find that the BH masses inferred by our MCMC for the 14 sources analysed occupy two well-separated regions (consistent with the single example above) corresponding to either low-mass BHs accreting above the Eddington limit, or very high-mass BHs accreting well below Eddington. In order to assess which solution is preferred by the MCMC analysis, we compare the mean likelihood associated with each solution and colour-code the data points according to the ratio between them. The colour coding therefore indicates which of the two solutions is more strongly represented in the overall posterior distribution, with redder colours supporting to the lower-mass solution. We clearly see that almost all sources prefer the low-mass solution, with only one object showing comparable likelihoods. At the same time, we find that the \citetalias{Maiolino2025} results struggle to explain most of the sources, as they lie between the two inferred solutions and are only partially consistent with the high-mass, less favoured one. However, for several objects the high-mass solutions imply black hole masses significantly larger than those expected from local correlations, by up to one order of magnitude. Such a result is physically unlikely as it would require an ADAF-like disc solution that the tables we employed do not model properly. It is also not clear whether at such low accretion rates the BLR would be present. Even more importantly, this solutions would imply unreasonably high BH-to-stellar mass ratios, predicting a BH mass comparable or even higher than the host stellar mass for a large fraction of the analysed systems.

A further noteworthy feature of the performed MCMC analysis is that the inclusion of the X-ray non-detection appears to be able to constrain the MBH spin more tightly than in \citetalias{Lupi2024}, where the broad-line observables alone were consistent with almost any spin value. This is shown in Fig. \ref{fig:corner}, for the case of \textsc{GS8083}, where the overall posterior spin distribution seems to prefer low spin values. We shall explore this aspect in more detail in Section~\ref{sec:spin}.

\subsubsection{Joint analysis of X-rays and broad-line emission}
\label{sec:MCMC_joint}
In the previous section, we constrained the best-fitting model parameters of the MCMC relying on the X-ray non-detection of the considered AGN sample, and on their estimated bolometric luminosities. 
It has to be highlighted, however, that the $L_{\rm bol}$ values provided for these object were not inferred by the continuum, which is often significantly contaminated by the host galaxy. Instead, their bolometric luminosity was estimated from the broad component of the $H\alpha$ emission line, relying on the standard scaling relation from \citet{Stern2012}. Here, there are two important caveats to consider: i) this scaling relation is characterised by a substantial intrinsic uncertainty, with a $1 \sigma$ scatter of approximately $\approx 0.4 ~\rm dex$; ii) at super-Eddington accretion rates, the radiative efficiency drops rapidly, while the collimation due to the funnel significantly alters the standard scaling between the continuum and the emission lines, as shown in \citetalias{Lupi2024}.

\begin{figure*}
\centering
\includegraphics[width=\hsize]{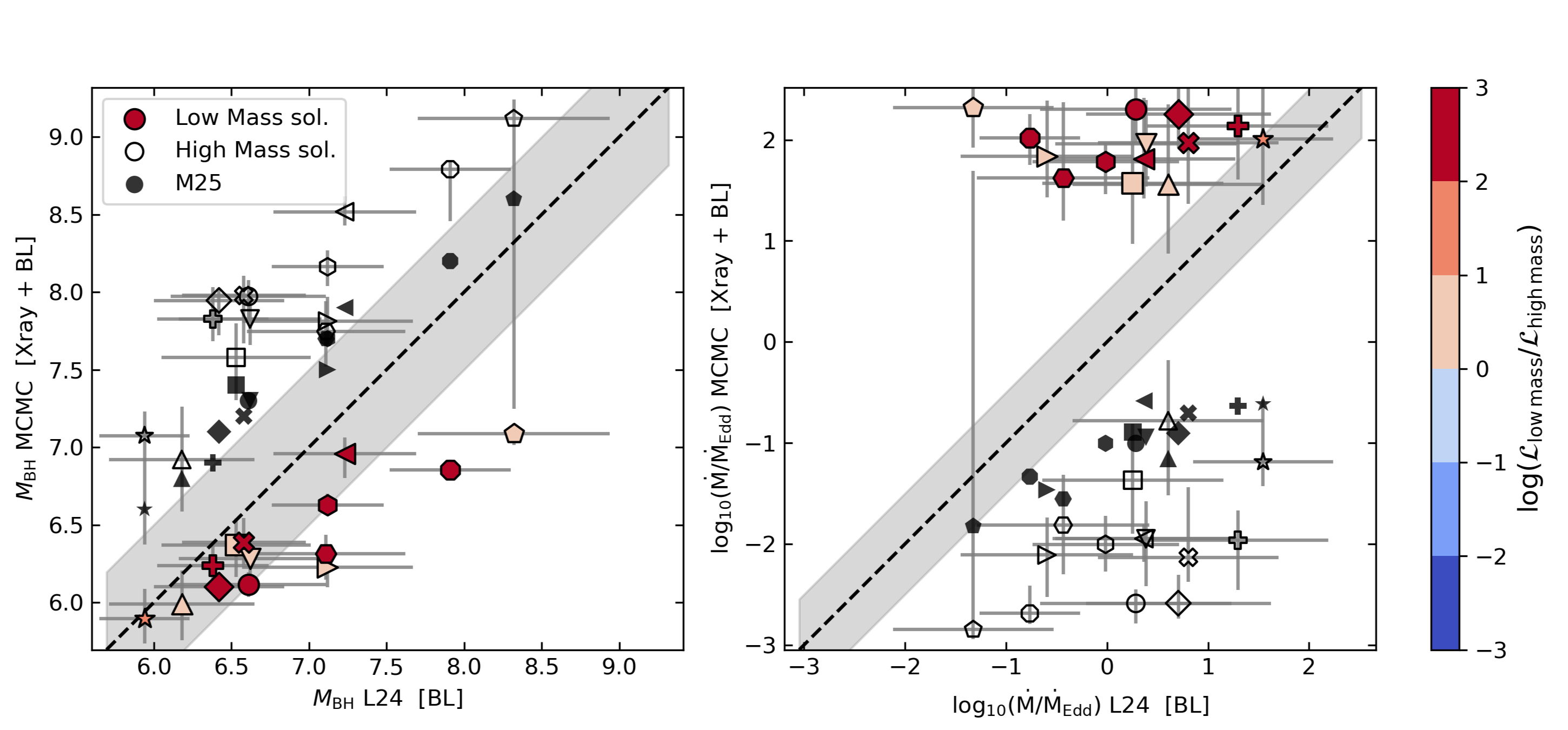}
\caption{Same as Fig.~\ref{fig:mcmc_vs_L24}, but here we compare the $M_{\rm BH}$ and $\dot{M}/\dot{M}_{\rm Edd}$  estimates obtained from the MCMC combined analysis that includes both the AGN X-ray non-detection and the broad-line emission constraints, with those from \citetalias{Lupi2024}, where the MCMC analysis was based only on the broad-line emission properties. For reference, black data points show, on the y axis, the values reported in the literature based on single-epoch estimates. For each object, the low-mass, high-mass, and literature estimates are shown with the same marker.}
\label{fig:mcmc_vs_L24}
\end{figure*}

For this reasons, we provide here a different parameter estimation by performing a joint analysis between the X-ray non-detections and the width and luminosity of the detected broad $H\alpha$ lines, without making use of bolometric luminosity estimates based on local scaling relations.
The results of this joint analysis are shown in Fig.~\ref{fig:mcmc_vs_L24}. As in Section \ref{sec:MCMC_xray}, we again found a characteristic bimodality in the posterior distribution of the inferred parameters, and proceeded to split the samples between lower-mass super-Eddington solutions and higher-mass sub-Eddington ones. 
The resulting best-fit values for $M_{\rm BH}$ and $\dot{M}/\dot{M}_{\rm Edd}$ from the joint MCMC analysis are plotted this time as a function of the estimates from \citetalias{Lupi2024}, in order to provide a direct comparison with the estimates obtained relying only on information from the observed broad emission lines.
We find that the preference for the low-mass solution is even more evident, as indicated by the ratio of the likelihoods, and no system seems to favour the high-mass one. As a reference, we also report in the plot the estimates by \citetalias{Maiolino2025}. Interestingly, the highly super-Eddington, low-mass solution from this work explains well the population with $M_{\rm BH} \lesssim  10^{7.5}\,\msun$ from \citetalias{Lupi2024}, with masses in line with (or even lower than) our previous estimates, whereas the low-Eddington, high-mass solution is again larger than the single-epoch estimates reported by \citetalias{Maiolino2025}. Note, also, that the Eddington ratio for this massive population is very low, sometimes close to $\dot{M}/\dot{M}_{\rm Edd}\simeq 10^{-3}$, for which the existence of the BLR is even more unlikely. 
In two of the analysed systems, for which the \citetalias{Lupi2024} analysis based only on broad lines suggested $M_{\rm BH} \gtrsim 10^8 ~\rm \msun$, the combined analysis prefers instead a significantly lower BH mass of $\approx 10^7 ~\msun$, and very high accretion rates $\log(\dot{M}/\dot{M}_{\rm Edd}) \gtrsim 2$. For the most massive of them, the corresponding high-mass solution from the MCMC is instead highly uncertain, and remains consistent within $1 \sigma$ with the \citetalias{Lupi2024} estimates.

\begin{figure}
\centering
\includegraphics[width=\hsize]{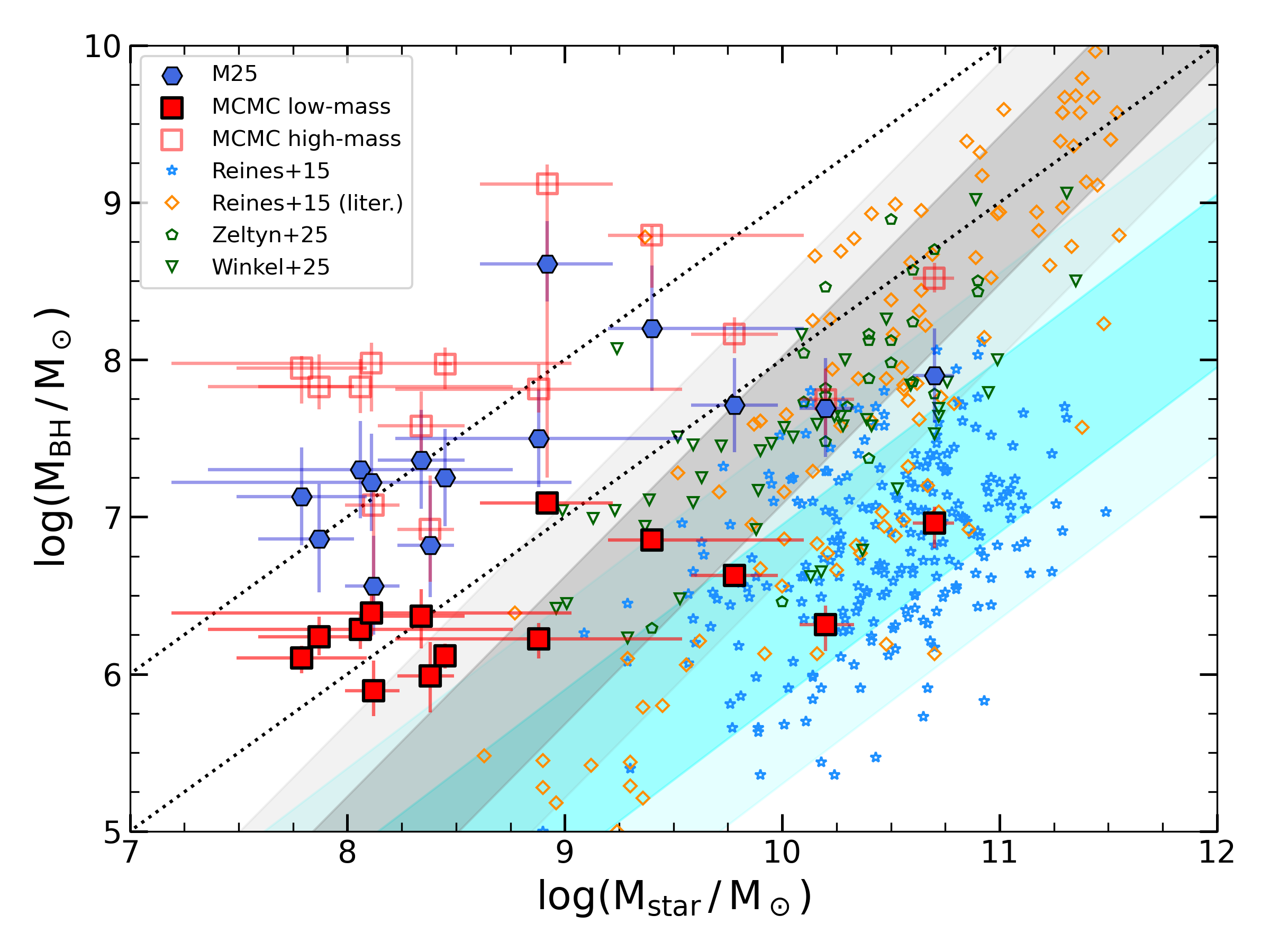}
\caption{Black hole vs stellar mass relation for the 14 sources in the analysed sample. Filled and empty red squares show the results from the combined MCMC analysis, assuming the low-mass and high-mass solution, respectively. Filled blue data points show instead the original black hole mass estimates from \citetalias{Maiolino2025}. The local AGN sample from \citet{reines2015} is shown with blue stars and orange diamonds, while the underlying shaded regions indicate the $1 ~\sigma$ and $2 ~\sigma$ uncertainties around the best-fit relations for local inactive (grey) and active (cyan) galaxies. Note that we also include local AGN observational data from \citet{winkel2025} and \citet{Zeltyn2025}, which are more consistent with the inactive galaxy distribution. For reference, dotted black lines mark constant black hole–to–stellar mass ratios of 0.01 and 0.1.}
\label{fig:Mbh_Mstar_MCMC}
\end{figure}

These results suggest that the combination of broad-line emission and lack of X-rays might be naturally reproduced by MBHs with relatively low masses, and hence much closer to the local scaling relations, accreting at very large Eddington ratios. To show this, we report in Fig.~\ref{fig:Mbh_Mstar_MCMC} the $M_{\rm BH}-M_{\rm star}$ relation for the low-mass solution we obtained, compared with local observations and the estimates by \citetalias{Maiolino2025}. We immediately note how the super-Eddington solution brings the BH mass on the local relation or very close to it, in agreement with what has been found in \citetalias{Lupi2024}. 
Conversely, the high-mass solutions provided by the MCMC would further exacerbate the tension with the local scaling relations, predicting most of the systems to have BH-to-stellar mass ratios as high as $\sim 0.1-1$.

\begin{figure}
\centering
\includegraphics[width=\linewidth]{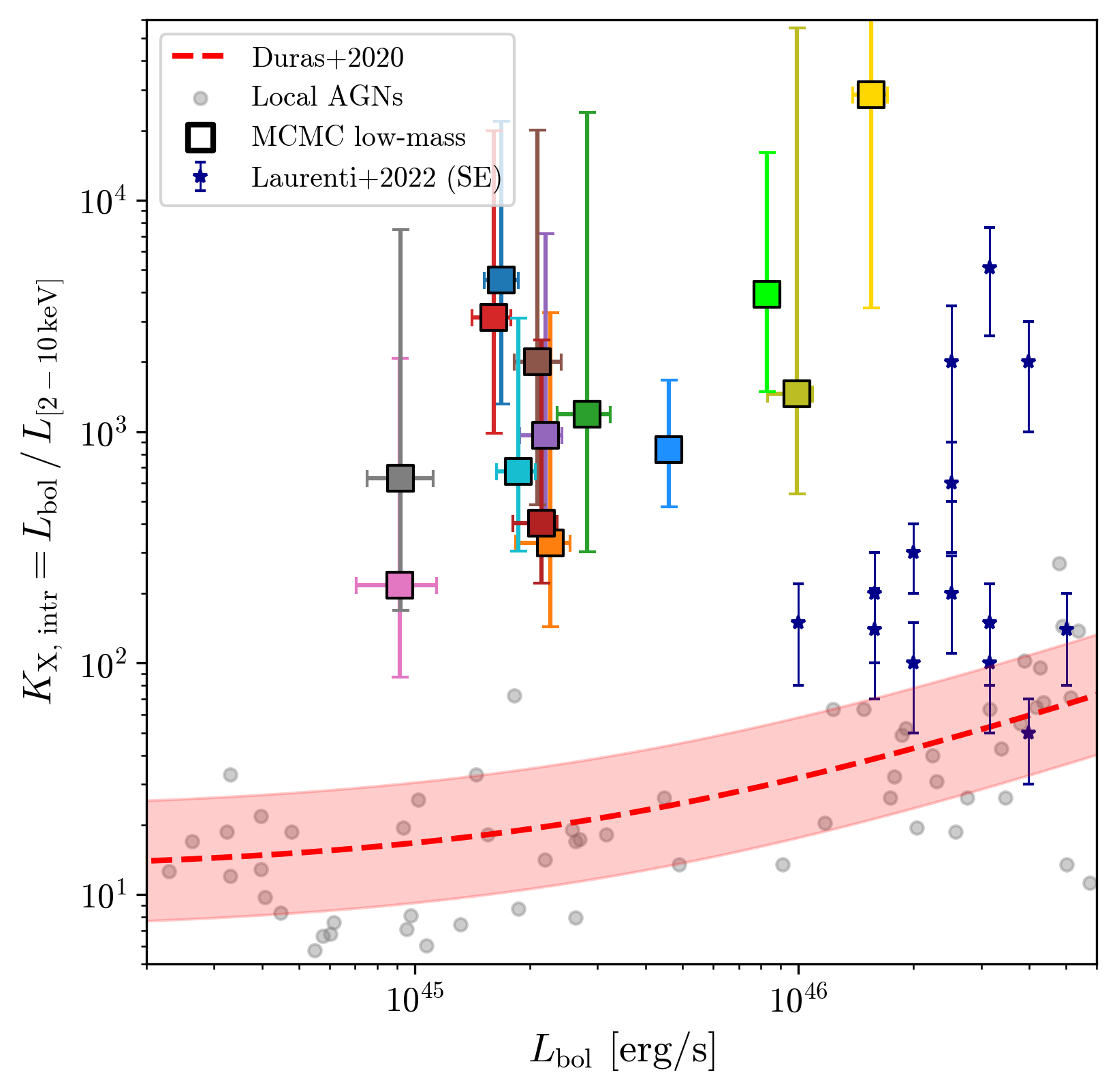}
\caption{Intrinsic bolometric correction in the restframe [2-10] keV energy range ($K_{\rm X, \, [2-10]\, keV}$) as a function of the AGN bolometric luminosity. Coloured data points show the estimates for the selected AGN population obtained from the combined MCMC analysis, assuming the low-mass solution shown in Fig. \ref{fig:mcmc_vs_L24}. As a reference, we report with blue data points the results from \citet{Laurenti2022} for a population of highly accreting AGNs ($\lambda_{\rm Edd} > 1$) observed at $0.4 \leq z \leq 0.75$. Despite being characterised by typical bolometric luminosities $\sim 1 ~\rm dex$ higher than our sample, these systems show a comparable intrinsic X-ray weakness.}
\label{fig:Kx_mcmc}
\end{figure}

Finally, in Fig.~\ref{fig:Kx_mcmc} we report the intrinsic bolometric correction inferred from the joint MCMC analysis, considering only the favoured, low-BH mass solution. We find that the predicted bolometric corrections range between $200 \lesssim k_{X,intr} \lesssim 3 \times 10^4$, displaying a mild increasing trend with bolometric luminosity. 
These values are significantly higher that those typically inferred for standard local AGNs, commonly described by the relation presented in \citet{duras2020}. However, when considering the $M_{\rm BH}$ and $f_{\rm Edd}$ estimates favoured by the MCMC posterior, it becomes clear that the highly accreting, super-Eddington sources dominating our sample have no direct counterparts in the local AGN population considered in \citet{duras2020}, in terms of either physical structure or emission properties. To provide a more meaningful comparison, we also report in Fig. \ref{fig:Kx_mcmc} the bolometric correction factors estimated in \citep{Laurenti2022} for a sample of 14 highly accreting ($\lambda_{\rm Edd} > 1$) AGNs at $0.4 \leq z \leq 0.75$, which are found to present a significant intrinsic X-ray weakness. Although this sample is characterised by $L_{\rm bol}$ $\approx 0.5-1$ dex higher than those of our high-redshift sources, their inferred bolometric corrections are broadly consistent with our estimates, ranging between $50 \leq k_{\rm X} \leq 5 \times 10^3$.

In contrast with the BH mass estimates adopted in the literature for the analysed JWST sample, the super-Eddington solutions favoured by our MCMC analysis imply intrinsically redder spectra, together with systematically higher bolometric luminosities. As a consequence, the corresponding data points are shifted towards higher luminosities relative to the estimates of \citetalias{Maiolino2025} (shown in Fig.~\ref{fig:Kx_M25}). Notably, this shift places our sources in a luminosity regime where the local AGN population also exhibits a mild increase in bolometric correction. 
For completeness, we also inferred the intrinsic values of the UV-to-X-ray spectral index that are predicted for our low-mass, high-Eddington AGN solutions, finding $\alpha_{\rm OX} \sim [-1.4 ~, -2.2]$. This is consistent with an extrapolation of the local  $\alpha_{\rm OX}$ versus $\lambda_{\rm Edd}$ relation found in \citet{lusso2010} to higher accretion rates, and matches the larger scatter towards more negative values (down to $\alpha_{\rm OX} \approx -2.3$) observed in local super-Eddington counterparts \citep{Laurenti2022}.

\subsection{Black hole spin distribution}
\label{sec:spin}
\begin{figure}
\centering
\includegraphics[width=\hsize]{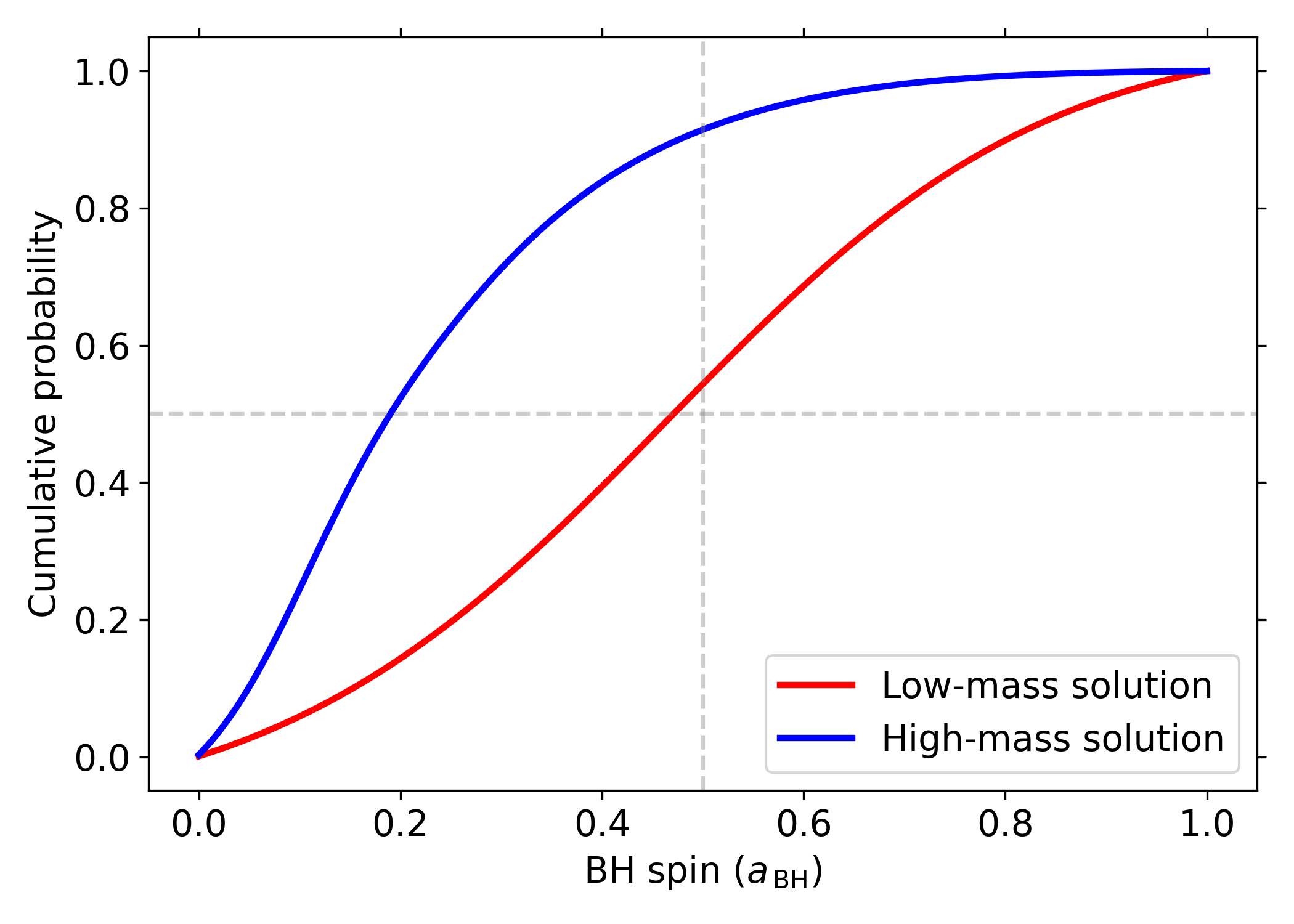}
\caption{Cumulative black hole spin distribution predicted by the combined MCMC analysis based on X-ray and broad-line emission. Red and blue curves show the results for the high-mass and low-mass solutions, respectively. Assuming the high BH mass solution, the analysed population would systematically require low spin values, with $a < 0.2$ for $\approx50 \%$ of sources, in order to remain consistent with the X-ray non-detections.
}
\label{fig:spin_distrib}
\end{figure}

In Figures \ref{fig:Kx_M25} and \ref{fig:Kx_L24} we showed that even for fixed values of BH mass and accretion rate, the expected X-ray bolometric correction would strongly depend on the BH spin. This could lead to differences of more than 1 dex in the inferred value of $K_X$. In the analysis presented in \citet{Lupi2024}, the BH spin values were only poorly constrained by the MCMC analysis, due to the limited observational data provided by the broad emission lines and to their moderate sensitivity to spin. We therefore investigated whether the inclusion of the additional constraints derived from the X-ray non-detections of the same sources can yield more informative insights into the determination of the BH spin. Relying on the results presented in Section \ref{sec:MCMC_joint}, which are based on the joint analysis of X-ray and broad-line emission, we assumed the posterior spin distribution obtained from the MCMC analysis for each individual source to be Gaussian. We then combined the resulting distributions for the entire AGN sample to derive the overall spin distribution for both the high- and low-mass BH solutions. The resulting cumulative distributions are shown in Figure~\ref{fig:spin_distrib}. We found that the two solutions point towards very distinct spin distributions. In the case of high-mass, low-accreting BHs, the spin distribution is strongly skewed towards low values. In particular, it requires that $\gtrsim 50 \%$ of the population have spin parameters of $a_\bullet < 0.2$. This is driven by the need for low radiative efficiencies in order to remain consistent with the X-ray non-detection of these systems.
Conversely, for low-mass BHs accreting at super-Eddington rates, the spin distribution is much less constrained, peaking at $a_\bullet \approx 0.5$, with a comparable fraction of systems having $a_\bullet \gtrless 0.5$. This reflects the fact that, given the high Eddington ratios characterising this solution, the large bolometric corrections shown in Fig.~\ref{fig:Kx_mcmc} typically lie well above the current X-ray lower limits, and are therefore only weakly constrained by the observations.

\section{Discussion and conclusions} 
\label{sec:conclusions}
In this work, we extended the \citetalias{Lupi2024} analysis of the broad-line emission from high-redshift MBHs discovered by JWST to incorporate also the lack of X-ray detections observed in many of these sources. Building on the model by \citetalias{Lupi2024}, in which the self-shadowing effect naturally arising in slim and thick accretion discs reduces the size of the BLR and biases virial MBH mass estimates, we have incorporated the over-cooling of the coronal plasma proposed by \citet{Madau2024} into our MCMC analysis tool. Our results suggest that the concurrent presence of broad Balmer line emission and extreme X-ray weakness that characterise the AGN population observed at $z>4$ with JWST can be explained by two qualitatively different populations: (i) a population of very massive, almost quiescent BHs, or (ii) a population of low-mass BHs accreting at highly super-Eddington rates. 

The first scenario, however, appears physically less plausible. On the one hand, it would imply $\rm M_{BH}/M_{\star}$ ratios even more extreme than those reported in the literature, posing a serious challenge to nearly all astrophysical models of MBH evolution; on the other hand, it would require relatively low MBH spin values, which are difficult to reconcile with such an early and rapid growth history, unless efficient spin-down episodes have previously occurred, which, however, are typically driven by strong jet emission during phases of super-Eddington accretion \citep{Narayan2022, Ricarte2023}. In addition, this high-mass solution is strongly disfavoured by our MCMC analysis in terms of integrated posterior probability.

The second scenario, involving super-Eddington accreting systems, would instead produce an intrinsically red spectrum, which may more naturally account for other observed properties of these sources \citep{Liu2025,Zucchi2026}. In this case, the inferred BH masses would be lower by $\sim 0.5 - 1.5 ~\rm dex$ compared to estimates based on single-epoch virial calibrations, bringing the systems closer to the local scaling relations, while still allowing for some intrinsic overmassiveness at the low-mass end, with $\rm M_{BH}/M_{star} \sim 0.01$. The highly accreting AGN population in this scenario would be characterised by steep X-ray spectra, implying bolometric correction factors of $K_{X} \sim 10^2 -10^4$, which, although significantly higher than those typical of the local AGN population, are consistent with what is observed in low-redshift samples of super-Eddington accreting systems \citep{Laurenti2022}.

An important caveat of our modelling is that the underlying accretion-flow structure is described using advective slim-disc solutions (through the \textit{agnslim} spectral model), rather than the geometrically thicker, non-advective radiation tori recently explored in \citet{Madau2026}. In such radiation-supported tori, photon trapping, multiple reflections within the funnel, and strong angular anisotropies in the emergent radiation field can differ significantly from the assumptions implicit in slim-disc spectra. As a consequence, our mapping between the accretion rate, funnel opening angle, X-ray bolometric corrections, and viewing-angle dependence should be regarded as pertaining to advection-dominated flows only. Non-advective thick-torus radiative-transfer calculations may modify the quantitative values of $K_{\rm X}$, $f_{\rm Edd}$, and the inferred spin distribution, although we expect the qualitative preference for low-mass, highly super-Eddington solutions to remain robust.

It should also be emphasised that our results assume the absence of extremely high gas column densities capable of absorbing the X-ray emission from the AGN. Recent studies focusing on the population of LRDs have shown that their peculiar spectral features may be linked to the presence of very dense gas surrounding the central source, such as strong Balmer absorption and pronounced Balmer breaks \citep{inayoshi2024,Deugenio2026, Ji2025, Naidu2025}, as well as potentially thermalised emission emerging in the rest-frame optical and near-infrared \citep{deGraaf2025}.
However, \citet{Liu2025} have recently shown that the same spectral features can be reproduced by the intrinsic continuum emission of super-Eddington accretion flows, in which opacity effects across the Balmer limit give rise to both a red optical spectrum and a Balmer break, without invoking additional absorbing material.
Nevertheless, since LRDs are excluded from our analysed AGN sample, we have not considered the contribution of dense absorbing gas in our modelling.

In addition, it is important to stress that LRDs represent only $\sim 10-30\%$ of the broad-line AGN population detected at high redshift. Consequently, the vast majority of sources still exhibit a complete lack of X-ray detection, without clear indications of dense gas along the line of sight. If, as has recently been suggested by \citet{Brazzini2026}, LRDs and the broader population of ``little blue dots'' are powered by the same central engines and differ primarily in geometry and/or in the evolutionary stage of their surrounding environments, this would imply that both populations are intrinsically X-ray weak. In this framework, the presence of dense gas in the minority of sources displaying a steep red optical continuum would further enhance their apparent X-ray weakness. 

Previous results from \citet{Sacchi2025} argued that softer, super-Eddington spectra alone would not be sufficient to explain the non-detection of stacked LRDs in \textit{Chandra} deep fields, assuming that the BH mass and bolometric luminosity estimates for these systems are accurate. We show, however, that the high-accretion solution naturally leads to significantly lower BH masses than currently inferred, which would fully alleviate the observed tension and could be further aided by partial gas obscuration.

Neglecting the presence of high gas column densities  in the analysed sample of BL AGNs nonetheless represents a potential caveat of our analysis, as it would bias our mass estimates. We note, however, that in the case of extremely large column densities capable of strongly attenuating the X-ray flux, specific gas configurations would be required to suppress the X-ray emission while allowing the ionising radiation responsible for the broad-line emission to remain largely unaffected \citep[see e.g.][]{Maiolino2024bhs, inayoshi2024}.
Although not explicitly considered in our analysis, several other physical effects may further increase the uncertainties of current BH mass estimates for this high-redshift population. These include electron scattering in dense gas environments \citep{Rusakov2025}, radiation beaming \citep{king2024}, and small-scale outflows, which might all contribute to the overall mass overestimation.

Future observations will provide crucial insights into the physical conditions and detailed spectral properties of this high-redshift AGN population. In particular, improved multiwavelength coverage and higher signal-to-noise spectroscopy will be needed to disentangle the relative importance of these different effects and establish more robust constraints on the properties of their central engines, especially in terms of BH masses and accretion rates.

\begin{acknowledgements}
AT, AL, and FH acknowledge support by the PRIN MUR “2022935STW" funded by European Union-Next Generation EU, Missione 4 Componente 2, CUP C53D23000950006. AT acknowledges financial support from the Bando Ricerca Fondamentale INAF 2023, Mini-grant ``Cosmic Archaeology with the first black hole seeds" (Ob.Fu. RSN1 1.05.23.04.01). 
\end{acknowledgements}

\bibliographystyle{aa}
\bibliography{aa59544-26}{}

\begin{appendix} 
\section{Corona properties}
\label{app:corona}

We have simplified the prescriptions of \citet{Madau2024} for super-Eddington flows as follows. We assume that a fraction $f_c$ of the available gravitational power is dissipated outside the optically and geometrically thick accretion flow in a hot corona. Magnetic buoyancy and fast reconnection have been postulated as possible mechanisms able to create a population of mildly relativistic, low-density electrons. We further assume that electrons in the corona follow a relativistic Maxwellian energy distribution. Under typical conditions the main cooling mechanism is unsaturated inverse Compton scattering.  
Under these model assumptions, we can write simple flux-balance equations  (cf, \citealt{Haardt1991}):
\begin{equation}
\begin{aligned}
F_s &= (1-f_c)\,F_g + (1-\epsilon/2)(1-a)F_c,\\
F_c &= f_c\,F_g + F_s,
\end{aligned}
\label{eq:balance1}
\end{equation}
where $F_g$ is the gravitationally generated radiation flux locally emitted from the disk photosphere, $\epsilon\equiv \Delta\Omega/2\pi$ is the normalised solid angle formed by the disc funnel (assumed to be conical for simplicity), and $a$ is the disc albedo. 

\begin{figure}
\centering
\includegraphics[width=\columnwidth,trim=13.cm 5.5cm 5.0cm 4.5cm,clip]{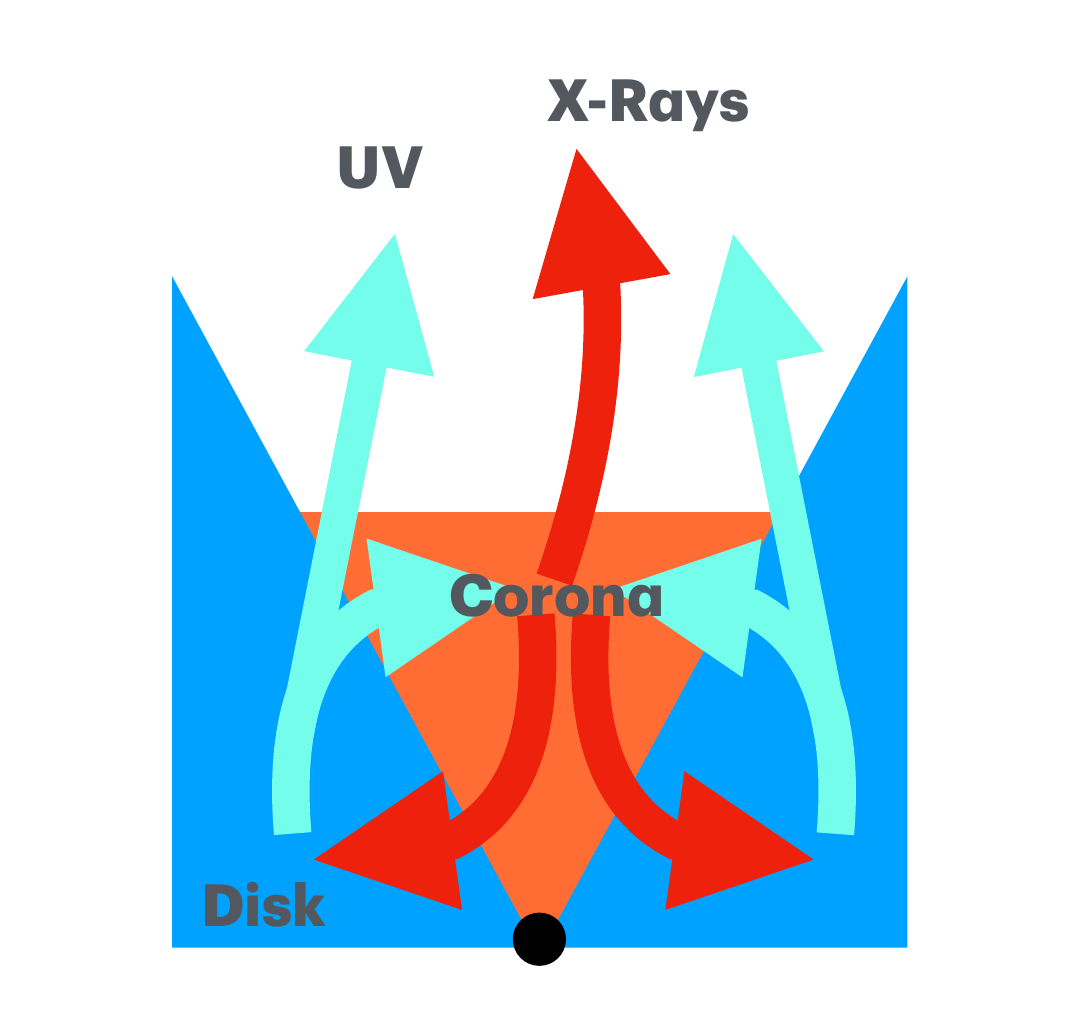}
\caption{Schematic illustration of the disk-corona radiative coupling in super-Eddington accretion flows. Soft UV photons from the optically thick disk enter a hot corona confined within the narrow funnel, where they are Compton upscattered by the coronal plasma. In a geometrically thick funnel, the corona is additionally exposed to incoming radiation and multiple reflections from the surrounding walls, enhancing Compton cooling and yielding softer X-ray spectra than in the standard thin-disk scenario \citep{Madau2024}.}
\label{fig:Cartoon}
\end{figure}

The theory of thermal Comptonisation links the Compton flux $F_c$ to the soft-photon input $F_s$ through 
\begin{equation}
    F_c = pA\,F_s + F_s,
\end{equation}
where $A\simeq 16\Theta^2+4\Theta$ is the inverse Compton amplification factor as a function of the dimensionless electron temperature of the corona $\Theta\equiv kT_e/m_ec^2$, and $p$ is the fraction of soft photons that are actually scattered by the corona. Soft photons can leave the funnel unscattered if the optical depth of the hot layer is $\tau\leq 1$. If the scattering probability in crossing once the hot-corona layer is $p_1$ (with $p_1\simeq \tau$ for $\tau\ll 1$), multiple reflections on the accretion-flow funnelled surface when $\epsilon<1$ increase the total scattering probability to 
\begin{equation}
    p = \frac{p_1}{1-(1-p_1)(1-\epsilon)}.
\end{equation}
Note that for $\epsilon=1$ (i.e., a standard plane-parallel disc) $p=p_1$, as expected. 
Finally, with the above set of relations, the amplification can be written as 
\begin{equation}
    A = \frac{1-(1-\epsilon/2)(1-a)}{p\big[(1-f_c)/f_c+(1-\epsilon/2)(1-a)\big]},
\end{equation}
and, for any given values of $f_c$, $p_1$, and $\epsilon$, the coronal equilibrium temperature $\Theta$ can be readily computed. A schematic illustration of the disk–corona configuration considered in this work is shown in Figure \ref{fig:Cartoon}.

\begin{figure}
\centering
\includegraphics[width=\columnwidth,trim=0cm 6cm 0cm 4cm,clip]{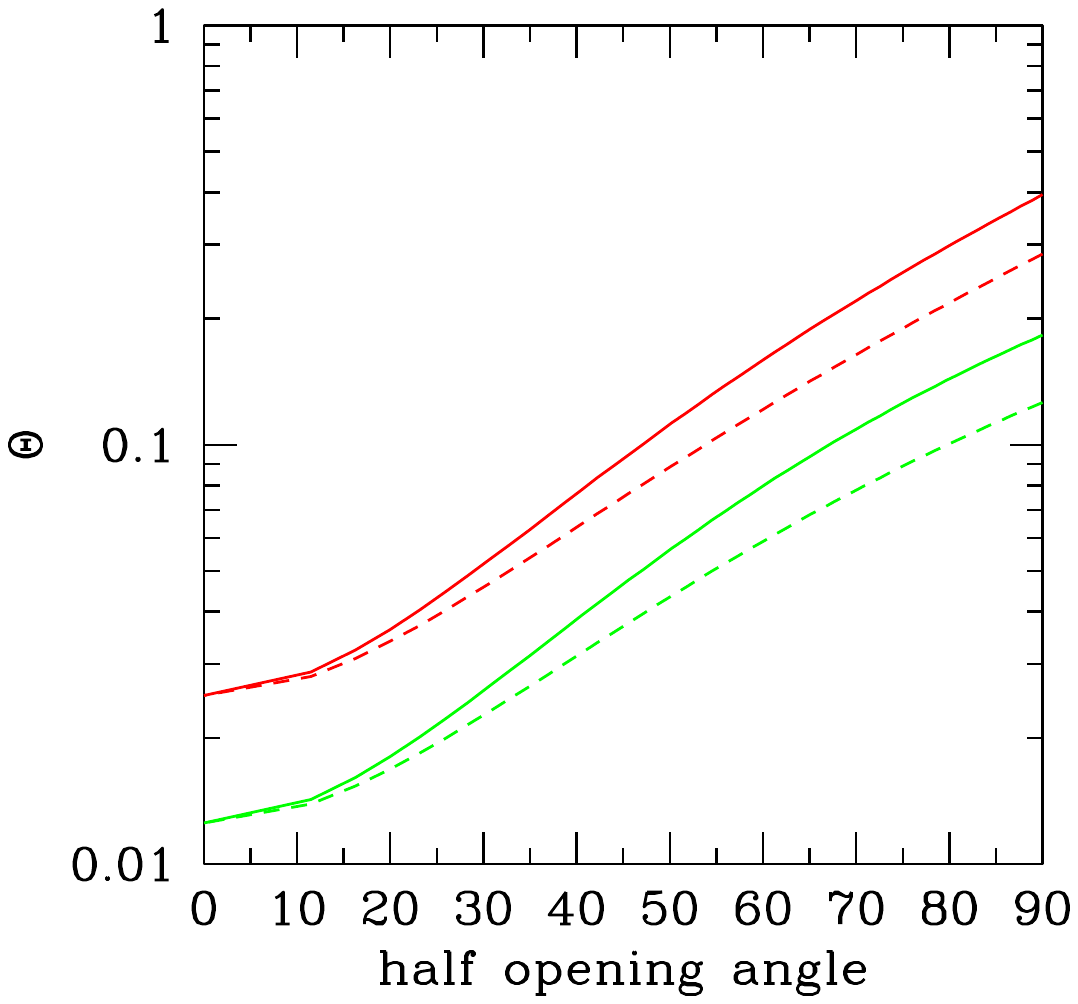}
\caption{Dimensionless corona temperature $\Theta\equiv kT_e/m_ec^2$ as a function of the funnel half-opening angle (in degrees). Red curves assume that all gravitational power is dissipated in the hot corona ($f_c=1$), while green curves correspond to $f_c=0.5$, with the remaining power released as thermal radiation in the accretion flow. Solid lines assume a single-pass scattering probability of soft photons by the corona of $30\%$, dashed lines of $50\%$.}
\label{fig:Theta}
\end{figure}

As an example, we show in Fig.~\ref{fig:Theta} the dependence of $\Theta$ on the funnel half-opening angle for two different values of $f_c$ (red vs green lines) and $p$ (solid vs dashed lines).

\end{appendix}

\clearpage

\end{document}